# Individual and Social Requirement Aspects of Sustainable eLearning Systems

*Ahmed D. Alharthi\* and Maria Spichkova*

School of Scince, RMIT University, Australia
\*Presenting author, E-mail: E-mail: ahmed.alharthi@rmit.edu.au

**Abstract:** Internationalization of the higher education has created the so-called *borderless university*, which provides better opportunities for learning and increases the human and social sustainability. eLearning systems are a special kind of software systems, developed to provide a platform for accessible teaching and learning, including also online access to learning materials and online support for learning and teaching. The aim of our current work is to extract, analyse, and combine the results from multiple studies in order to develop an RE framework for sustainable eLearning systems.

We call a system sustainable, if it has a positive effect on and whose direct and indirect negative impacts resulting from its development, deployment, and usage are minimal. Sustainability has various dimensions. We classify sustainability requirements of eLearning system to five dimensions: individual (human), social, technical, environmental, and economic. In this paper, we focus on human and social aspects (i.e., individual needs the relationship of people within society), as the eLearning systems have a very strong impact on human dimension of sustainability, where their impact on environmental dimension is rather small. This also provides us a basis to identify the corresponding requirements for sustainable eLearning systems. These requirements include collaboration, learner-centred features, leadership development and the reuse of the learning materials. As a result, achieving individual and social requirements for eLearning systems would provide higher quality of leaning and teaching, as well as better opportunities for learning and increasing the human and social sustainability.

**Keywords:** eLearning Systems; Sustainability; Learning; Teaching; Requirements Engineering.

## 1. Introduction

Rapid changes in society and technology demand that everyone could gain and update their knowledge and skills in distance education. Formal classes have been partially replaced or augmented by self-directed learning and flipped classrooms. According to Rahanu et al. (2015), over the last 30 years, teacher-centred approach has been shifted to a learner-centred approach because of the development of information and communication technologies and the social media revolution.

An eLearning system can be defined as an educational solution to deliver knowledge, facilitate learning and improve performance by creating, using and managing appropriate technological processes and resources, cf. Ghirardini (2011) and Richey (2008). One popular example of eLearning system is Learning Management System (LMS) that includes virtual classroom, collaboration functions, and instructor-led courses. As per Dagger et al. (2007), an LMS has two types:
- Proprietary LMS, e.g., Blackboard and Desire2Learn, and
- Open-source LMS, e.g., Moodle and Sakai.

Naumann et al. (2011) define sustainable software as "*software, whose direct and indirect negative impacts on economy, society, human beings, and environment that result from development, deployment, and usage of the software are minimal and/or which has a positive effect on sustainable development*". Robertson (2008) defines sustainable e-learning as "*e-learning that has become normative in meeting the needs of the present and future*" and he used active theory in his study to describe when organisational, technological and pedagogic activity systems come into contact to achieve sustainability. In our approach we follow these definitions.

Software sustainability has various dimensions. For example, Goodland (2002) suggested to consider *individual (human), social, economic, and environmental sustainability* dimensions. Penzenstadler and Femmer (2013) as well as Razavian et al. (2014) added to these dimensions a new one: *technical sustainability dimension*. Requirements engineering (RE), i.e., requirements elicitation, evaluation, specification, and design producing the functional and non-functional requirements, is one of the key disciplines in software engineering, as requirements-related errors are often a major cause of the delays in the product delivery and development costs overruns, cf. van Lamsweerde (2008). A number of studies showed that if a software system is developed without taking into account sustainability requirements, this system could have negative impacts on individual, social, technology, economic, and environment sustainability, cf. Berkhout and Hertin (2001), Lago and Jansen (2011), Naumann et al. (2011), Penzenstadler and Femmer (2013), Stepanyan et al. (2013). This is especially important for eLearning systems, as they deal not only with a large amount of teaching data, but also with a large number of users with diverse backgrounds (educational as well as cultural).

In our previous work (Alharthi et al, 2015) we introduced a general idea of an RE framework for eLearning systems, with the focus on users' diversity in background, culture, and regulations. The goal of the framework is to contribute to the RE process for development and improvement of ELearning systems, which might improve the overall sustainability of online and on-campus teaching and learning activities.

**Contributions:** The aim of our current work is to extract, analyse, and combine the results from multiple studies in order to develop an RE framework for sustainable eLearning systems. This paper provides results from multiple studies extracted from Systematic Literature Review (SLR) of sustainable eLearning system.

**Outline:** The rest of the paper is organised as follows. Section 2 presents the background for our research as well as related work. Section 3 provides analysis of individual and social sustainability requirements for eLearning systems. Section 4 concludes the paper, discussing the core contributing of the paper and future work.

## 2. Background and Related Work

There are several studies focusing on sustainability of eLearning systems. Robertson (2008) proposed in his study a notion of activity theory, and explained when organisational, technological and pedagogic activity systems cooperate to achieve increase sustainability by involving eLearning systems. Stepanyan et al. (2013) reviewed 46 papers limited to publications between 2000 and 2010, and mapped the area of sustainable e-learning three categorise having resource management, educational attainment and professional development and innovation. However, their studies covered individual, social and economic dimensions of sustainability, leaving out of scope technical and environmental dimensions.

To enable eLearning to be sustainable, Stewart and Khare (2015) analysed eLearning with respect of ecology, economy, culture, and politics domains and applied the Sustainability Circle Framework that developed by the Global Compact Cities Programme for urban sustainability profile of a particular city or region. This profile has four domains including ecology, economy, culture and politics. There are 7 sub-domains in each main domain in order to assist in assessment through the completion of a survey having 7 questions. The assessment is conducted on a nine-point scale that ranges from 1 being critical to 9 labelled vibrant. This method, which the authors presented, generates a clear graphic representation of the sustainability profile for eLearning systems. However, this framework needs to be reformulated to fit eLearning development. For instance, collaboration, which is part of individual dimension, is not included. Also, sustainability requirements may identify and follow sustainable software engineering in order to cover all the five dimensions and to be standardised with other software domains.

### 3. Individual and Social Sustainability Requirements for eLearning System

We conducted SLR of sustainability requirements for eLearning system that determined 15 sustainability requirements from 51 studies limited between 2005 and 2015 and then we classified them to five dimensions of sustainability requirements including individual, social, technical, environmental and economic dimensions. As a result of the SLR, 66% aspects are related to human dimension (individual and social) that we will focus on in this paper.

In the SLR, we followed the dimension differentiations defined by Goodland (2002), Penzenstadler (2014), Razavian et al. (2014):

- **Individual (human) sustainability:** Individual needs should be protected and supported in dignity and in a way that developments should improve the quality of human life and not threaten human beings;
- **Social sustainability:** Relationship of people within society should be equitable, diverse, connected and democratic;
- **Technical sustainability:** Technology has to cope with changes and evolution in a fair manner of respecting natural resources;
- **Environmental sustainability:** Natural resources have to be protected from human needs and wastes; and
- **Economic sustainability:** A positive economic value and capital should be ensured and preserved.

As presented in Table 1, we distinguish between two types of sustainability requirements: *general* (applicable to other domains, e.g., health systems domain) and eLearning system *specific*. In eHealth services, for example, personalisation feature is essential and assist to improve eHealth services Hine et al. (2008). On other hand, learner-centred features, reuse of the learning materials, learning object repository belong to education domain only, and should be seen as a specific requirements (features) of eLearning systems. Our study has shown that technical, environmental and economic are general sustainability requirements, as they could be identified and analysed for any kind of software.

Furthermore, as result of SRL, we found 34 studies (out of 51 studies we analysed) on the individual and social sustainability requirements of eLearning systems. We classified these studies into three types:

- **Empirical studies:** Knowledge is gained by observations or experience methods. As per Perry et al. (2000), an empirical study is a test comparing what we believe to what we observe in order to help us understand how and what things work;
- **Theoretical or conceptual studies:** Methods consisting of concepts with definition of knowledge being considered to describing a phenomenon of interest[1]; and
- **Hybrid studies:** Combinations of empirical and theoretical studies or other studies such as systematic reviews.

Table 1 Kind of Requirements of Sustainability Requirements of eLearning systems

| Dimension | Sustainability Requirements | Type of Requirement |
|---|---|---|
| Individual and Social | Personalisation | General |
| | Learner-Centered Features and Lifelong Learning | **Specific** |
| | Collaboration | General |
| | Leadership Development | General |
| | Privacy and Security | General |
| | Reuse of the Learning Materials | **Specific** |
| Technical | Learning Object Repository (LOR) | **Specific** |
| | Support of Shared Services | General |
| | Software Quality Requirements e.g. flexibility, and integrability. | General |
| | Portability | General |
| | Modularity | General |
| Environmental | Green and sustainable software engineering | General |
| | Cloud computing | General |
| Economic | Reducing the Budget | General |
| | Ensuring the Growth | General |

Figure 1 shows the classification result for the 34 studies on the individual and social sustainability requirements of eLearning systems. The 47% of the studies were classified as empirical study, while 44% of the studies are theoretical, and 9% have hybrid nature.

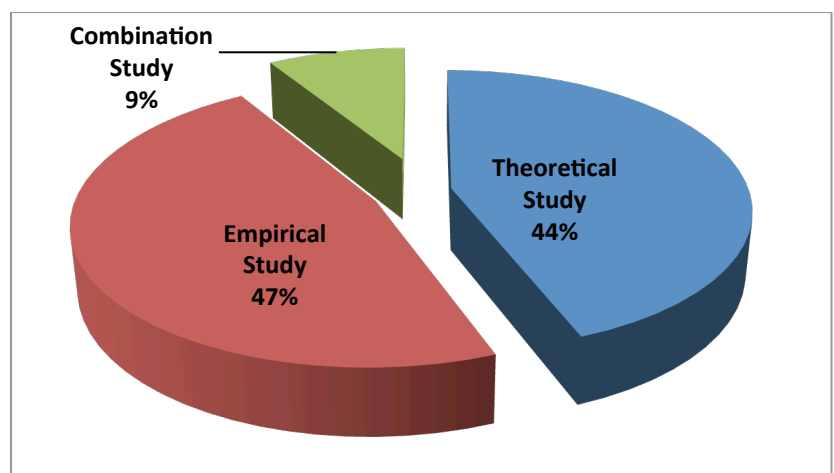

Figure 1 The Classifications of Studies in Percentage

---
[1] http://libguides.usc.edu/writingguide/theoreticalframework

A few studies from the empirical category, presented are well-structured and well-presented statistical data. For example, Louhevaara (2013) pointed out the background of participants such as their academic level, gender and age in their study that describing the characteristics of learning programs to promote sustainable well-being at work. On the other hand, some studies did not state the background or academic levels of their participants. For instance, Mridha et al. (2013) claimed that eLearning increases educational equity and English language proficiency has improved but they didn't show how much the increase and the improvement.

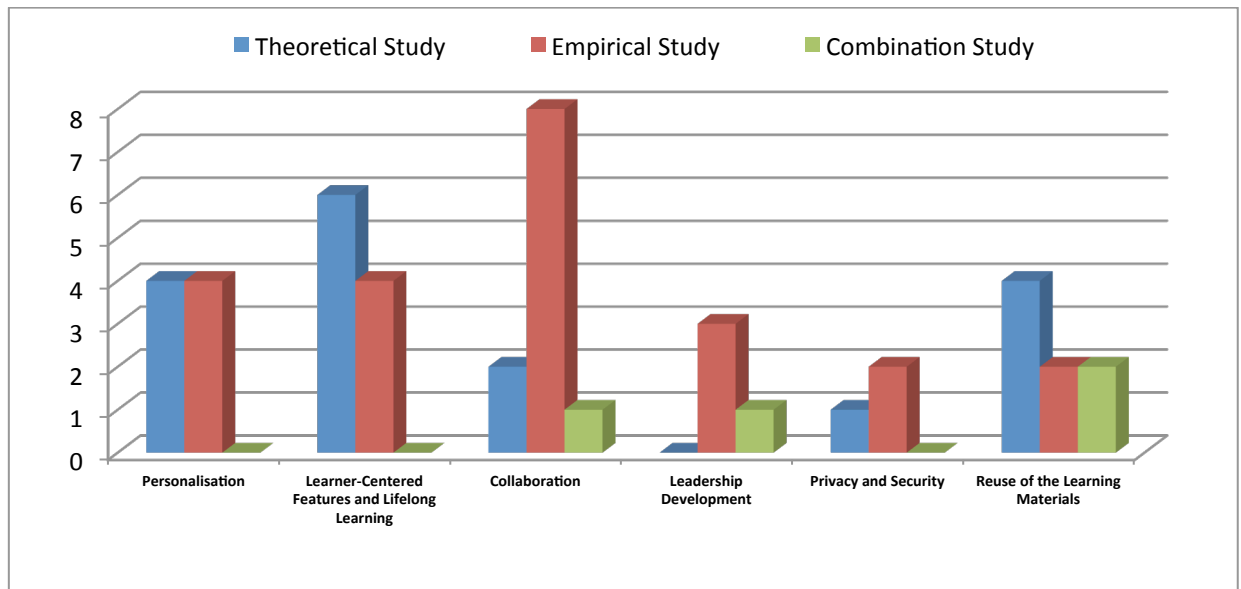

Figure 2 The Classifications of Studies for Each Requirement

Figure 2 illustrates in which study categories the particular individual and social sustainability requirements (listed in Table 1) were identified in the 34 studies. The *Personalisation* requirement was investigated almost both by theoretical and empirical studies in equal proportions, where the *Collaboration* and *Leadership development* requirements was dominantly investigated by using empirical methods, and the *Learner-centred features and life-long learning* and *Reuse of the learning materials* requirements were studied dominantly through theoretical methods. In what follows, we review the studies consistent on the individual and social sustainability requirements for eLearning systems in more details.

**3.1 Personalisation**
Ros et al. (2013) built a personal learning environments within iGoogle and conducted a survey based on a 5-point scale for 11 questions that has been responded by 150 students from post-graduate and graduate programs in the Faculty of Computer Science, Psychology, and Law. The authors concluded that there is need of the reference models which will let users discover services. Also, virtual laboratory was implemented with Moodle eLearning system to support free-open resources and personalisation Meneses (2011). Both studies highlighted the demand of personalisation requirements.

**3.2 Learner-Centered Features and Lifelong Learning**
Virtual world (OpenSim, an open simulator) was integrated with SLOODLE and Moodle environments by Pellas (2014). 94 students used the system and answered a survey having 38 questions. Pellas (2014) proposed that the use of virtual world could increase user's learning ability.

and Panetsos et al. (2008) explored the idea of lifelong on personal learning environments and academic library activity. The studies highlighted that eLearning systems require ability to be integrated with personal learning environment and support learner-centred and lifelong learning.

### 3.3 Collaboration
Ossiannilsson and Landgren (2012) used three benchmarking (E-xcellence+, the eLearning Benchmarking Exercise, and the e-learning quality model,) during two years in higher education in order to specify the critical success areas of eLearning. Similarly, Sridharan et al. (2010) evaluated the critical success factors of sustainable eLearning. Thus, collaborative technology is one of critical success factors of sustainable eLearning systems along with clear understanding of pedagogical theory of collaborative-learning.

### 3.4 Leadership Development
Several studies identified leadership development as a requirement of sustainable eLearning systems. Stepanyan et al. (2013) explored the sustainability of eLearning system and identified professional development and innovation one of three pillars of sustainable eLearning systems. The authors stated that a commitment of continuous development would benefit the adaption of change such as instructors training and educational leadership. Konting (2012) also mentioned that there is a need to improve young academic leaders to sustain eLearning system.

### 3.5 Privacy and Security
Roy (2012) as well as Stewart and Khare (2015) highlighted that privacy and security aspects of eLearning systems that require significant research. Pardo et al. (2012) proposed authoring system that supporting collaboration, easy re-purposing, and continuous updates. Thus, sustainable eLearning system should protect users' information and right, and provide secure environment.

### 3.6 Reuse of the Learning Materials
Vovides et al. (2014) described a study where five schools participated in implementing eLearning system to enhance the quality of the training by accessing and reusing digital resources in the Medical Education. The study of Luyt (2015) has shown that it requires eight to twelve months when instructors design their course with help of instructional designer. Therefore, the reuse of learning materials will enable designers to provide courses quickly and easily.

## 4. Conclusion
This paper presents core results of analysis of sustainability requirements for eLearning system. The analysis was based on the systematic literature review, which included 51 papers. In our current work, we focused on the individual and social sustainability requirements of eLearning systems, described in 34 studies (out of 51 studies we analysed). We classified these studies into three types: empirical, theoretical and hybrid studies, and analysed by what type of study the particular individual and social sustainability requirements were investigated by the authors.

In our future work we are going to conduct a survey on eLearning systems currently used in Australian and Saudi Arabian universities, to develop a sustainability profile framework for sustainable eLearning systems.

**Acknowledgements**

The first author is supported by a scholarship from Umm Al-Qura University, Saudi Arabia.